\documentclass[a4paper,11pt]{article}    
\usepackage{pos}
\usepackage{graphicx}
\usepackage{amsmath}
\usepackage[font=footnotesize]{caption} 

\title{Neutrino-Nucleus Scattering Simulation Using GENIE Event Generator}
\ShortTitle{Impact of Axial Mass in Neutrino Scattering for DUNE}

\author*[a]{Luiz Felipe Brandão Magalhães Rodrigues}
\author[a]{Gustavo do Amaral Valdiviesso}
\author[a]{Fernando G. Gardim}

\affiliation[a]{Instituto de Ci\^encia e Tecnologia, Universidade Federal de Alfenas, 37715-400 Po\c cos de Caldas, MG, Brazil}

\emailAdd{luiz.magalhaes@sou.unifal-mg.edu.br}

\abstract{The Deep Underground Neutrino Experiment (DUNE) relies on the precise reconstruction of incident neutrino energy to extract oscillation parameters. The use of Liquid Argon ($^{40}$Ar) introduces significant theoretical complexities due to nuclear effects. This work evaluates neutrino-nucleus scattering models using the GENIE Monte Carlo event generator. We demonstrate the limitations of the Relativistic Fermi Gas (RFG) model in the Charged Current Quasi-Elastic (CCQE) channel and its reliance on an inflated Axial Mass ($M_A$) to match experimental data from MicroBooNE. We show that transitioning to advanced many-body frameworks, specifically the Local Fermi Gas (LFG) supplemented with Random Phase Approximation (RPA), is needed. Additionally, we evaluate the impact of the 2021 Bodek-Yang parameter updates on Deep Inelastic Scattering (DIS), concluding that modernized constraints are a prerequisite for mitigating systematic biases in DUNE.}

\FullConference{
  The 43rd International Conference on High Energy Physics (ICHEP 2026)\\
}

\begin{document}
\maketitle

\section{Introduction}
The Deep Underground Neutrino Experiment (DUNE) is a next-generation project designed to measure neutrino oscillation parameters with unprecedented precision \cite{abi2020deep}. Because the neutrino energy is unknown, it must be reconstructed from the measured kinematics of the outgoing particles \cite{alvarez2018nustec}. This energy reconstruction is distorted by the nuclear environment of the $^{40}$Ar target \cite{alvarez2018nustec}. To accurately test the models' predictions, DUNE relies on Monte Carlo event generators, prominently the GENIE framework \cite{alvarez2018nustec}. This work, which is an upcoming master's dissertation \cite{RodriguesLFBsystematics}, investigates the phenomenologies of neutrino-nucleus scattering in GENIE, quantifying the limitations of the GENIE models and the necessity of advanced nuclear frameworks for DUNE, utilizing the SDumont supercomputer (LNCC).

\section{The Axial Mass and MicroBooNE Data}
\begin{figure}
    \centering
    \begin{minipage}{0.45\textwidth}
        \centering
        \includegraphics[width=\linewidth]{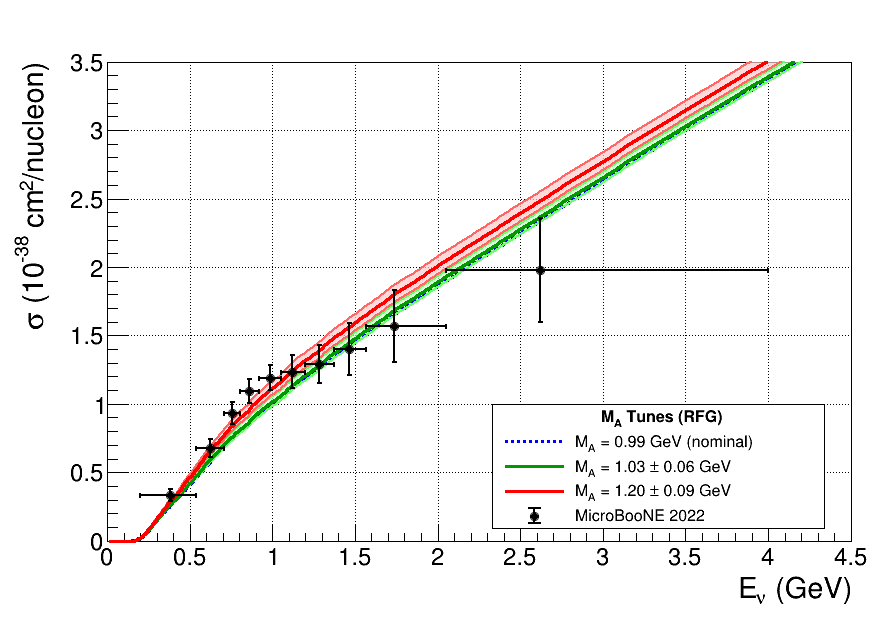}
        \caption{GENIE simulation of the $\nu_\mu$ inclusive cross-section on Argon compared to MicroBooNE (2022) data \cite{abratenko2022first} (black points). The curves represent the GENIE value evaluated at $M_A = 0.99$ GeV (blue dashed), $1.03$ GeV (green solid) from ref. \cite{abe2023updated}, and $1.20$ GeV (red solid) from ref. \cite{gran2006measurement}.}
        \label{fig:microboone}
    \end{minipage}
    \hfill
    \begin{minipage}{0.43\textwidth}
        \centering
        \includegraphics[width=\linewidth]{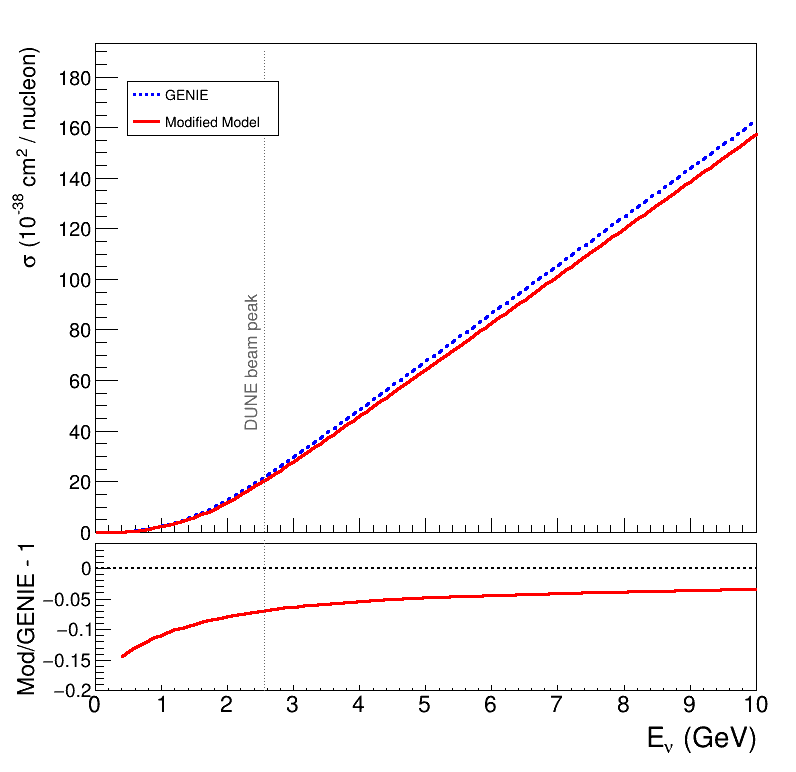}
        \caption{DIS cross-section comparison between GENIE and 2021 Bodek-Yang parameter updates from ref. \cite{bodek2021inelastic}. The bottom panel is the ratio, representing the percentual difference between the models - 1.}
        \label{fig:DIS_Comparison}
    \end{minipage}
\end{figure}

In the Charged Current Quasi-Elastic (CCQE) channel, the dipole approach of the axial form factor $F_A(Q^2)$ depends on the axial mass ($M_A$) which is an observable \cite{alvarez2018nustec}. While free nucleon data established $M_A \approx 1.03$ GeV \cite{abe2023updated}, modern experiments utilizing heavy nuclear targets found a severe underprediction of the total cross-section when applying the standard Relativistic Fermi Gas (RFG) model \cite{abratenko2022first}. To reconcile theoretical predictions with experimental data, the axial mass was inflated to an effective value of $M_A = 1.20 \pm 0.09$ GeV \cite{gran2006measurement}. To contextualize this parameter within Liquid Argon targets, we simulated the RFG predictions in GENIE and compared them against the inclusive $\nu_\mu$ cross-section data measured by MicroBooNE in 2022 \cite{abratenko2022first}. As shown in Fig. \ref{fig:microboone}, the experimental data align closely with the $M_A = 1.20$ GeV. However, inflating the axial mass to recover the event rate masks incomplete nuclear modeling \cite{alvarez2018nustec}. It is necessary to implement models that better describe nuclear dynamics. The Local Fermi Gas (LFG) model improves upon the RFG by accounting for the radial density of the nucleus via a Woods-Saxon distribution, providing a more physically realistic initial momentum distribution \cite{alvarez2018nustec}. 

\section{Deep Inelastic Scattering Updates}
Deep Inelastic Scattering (DIS) constitutes the largest fraction of the DUNE flux \cite{abi2020deep}. The GENIE framework utilizes the Bodek-Yang model. We evaluated the implementation of the modernized 2021 Bodek-Yang parameters \cite{bodek2021inelastic} against the GENIE tune. As shown in Fig. \ref{fig:DIS_Comparison}, the updated coefficients induce a suppression of the DIS cross-section, corresponding to an approximately 7\% reduction at the DUNE energy peak (2.5 GeV) \cite{abi2020deep}.

\section{Conclusion}
\cite{RodriguesLFBsystematics}The systematic evaluation using the GENIE event generator demonstrates that the standard Relativistic Fermi Gas model is insufficient for high-precision oscillation physics. Relying on parameter adjustments to reconcile theoretical predictions with Liquid Argon data can severely bias DUNE's energy reconstruction. Instead, we need more robust models accounting for more precise nuclear dynamics. The update of the modernized constraints for DIS is fundamental for mitigating systematic backgrounds and ensuring the success of DUNE's measurements.

\bibliographystyle{unsrt}
\bibliography{Bibliography}

\end{document}